\documentclass[%
 aps,
 prb,%
 amsmath,amssymb,
 preprint,%
]{revtex4-1}

\usepackage{color}
\usepackage{graphicx}
\usepackage{hyperref}
\usepackage{dcolumn}
\usepackage{bm}

\draft 

\begin{document}
\title{{Tuning the Conductance of Monatomic Carbon Chain}}

\author{Xi Chen}
\affiliation{Institute of Modern Physics, Fudan University, Shanghai 200433, China}
\affiliation{Applied Ion Beam Physics Laboratory, Key Laboratory of the Ministry of Education, Fudan University, Shanghai 200433, China}

\author{Chen Ming}
\affiliation{Institute of Modern Physics, Fudan University, Shanghai 200433, China}
\affiliation{Applied Ion Beam Physics Laboratory, Key Laboratory of the Ministry of Education, Fudan University, Shanghai 200433, China}

\author{Fan-Xin Meng}
\affiliation{Institute of Modern Physics, Fudan University, Shanghai 200433, China}
\affiliation{Applied Ion Beam Physics Laboratory, Key Laboratory of the Ministry of Education, Fudan University, Shanghai 200433, China}

\author{Jing-Tian Li}
\affiliation{Institute of Modern Physics, Fudan University, Shanghai 200433, China}
\affiliation{Applied Ion Beam Physics Laboratory, Key Laboratory of the Ministry of Education, Fudan University, Shanghai 200433, China}

\author{Jun Zhuang}
\affiliation{Department of Optical Science and Engineering, Fudan University, Shanghai 200433, China}

\author{Xi-Jing Ning}
\email[]{Electronic mail: xjning@fudan.edu.cn}
\affiliation{Institute of Modern Physics, Fudan University, Shanghai 200433, China}
\affiliation{Applied Ion Beam Physics Laboratory, Key Laboratory of the Ministry of Education, Fudan University, Shanghai 200433, China}

\date{\today}

\begin{abstract}
  \emph{Ab initio} calculations show that the conductance of short monatomic carbon chain can be dramatically modified by adhering a single H, N, or O atom to the chain. For example, the conductance of the pristine chain gets about two orders of magnitude smaller if an H atom is adhered to the chain. By a statistical model, the structure of the carbon chain with the single atom adhered is found to be quite stable at room temperature, indicating that the method can be used to tune the conductance of monatomic carbon chain.
\end{abstract}


\maketitle

\section{Introduction}
Searching for nano-materials to miniaturize the electronic devices has been a highly concerned topic during the last decades. Owning the true 1D structure as the ultra-thin nanowire, monatomic carbon chain (MCC) has captured attentions since the concept was raised up. In 2007, a method of pulling monatomic chains out of a graphene sheet was theoretically proposed\cite{12}, and was implemented in the same year by an experiment\cite{13}. In 2009, stable and rigid carbon atomic chains were unambiguously observed by removing carbon atoms from graphene through the controlled energetic electron irradiation inside a transmission electron microscope\cite{14}. These experiments triggered intense theoretical studies on MCCs. In fact, even before the evident characterizations, the MCCs had been used as models for years to calculate the electrical and optical properties of atomic chains. Lang and Avouris \cite{16} found that even-numbered and odd-numbered carbon chains (cumulenes) have different conductance, which might be caused by the difference in the density of states (DOS) at the Fermi level of the combined electrode-atomic-wire system of the free chains. Crljen et al\cite{15} investigated the transport property of another form of carbon atom chain, polyynes with metal electrodes and its dependence on the length of the molecular chain. Additionally, Nyk\"anen et al investigated the transport properties of hydrogen capped MCCs, polyynes and cumulenes, on gold and silver surfaces\cite{17}, suggesting that the C-C bond and the density of states at the Fermi level of the entire system might play an important role in the conductance.

However, there are few studies addressing practical ways to modulate the conductance of MCCs up to now. In the present work, we suggest modifying the conductance by adhering an atom or molecule to the chain. Thus, we should firstly answer if the structure is stable at room temperature for practical uses, although the pristine MCC has been predicted to be very stable\cite{21}. Since the gas molecules of H$_2$, N$_2$, and O$_2$ are the most common components of the atmosphere, they should be considered firstly. But the results show that these molecules cannot be adhered to MCCs stably at room temperature. Therefore individual atoms of these gases (i.e. H, N, and O atom) were used instead. Firstly the stability of MCCs with an adhered atom was examined by a statistical model\cite{21}, which has been successfully used in several aspects such as the characteristics of nano-optoelectronic devices e.g. stability\cite{21,22,23,24,25}, the growth probability prediction of isomer clusters\cite{26}, and the prediction of thermal chemical reaction rate\cite{27}. Then the transport properties of MCCs with an H, N or O atom adhered to different sites were investigated.

\section{Models and Methods}

\subsection{The statistical model}

The statistical model is based on the fact that the kinetic energy ($\varepsilon$) of a single atom in condensed matters obeys the Boltzmann distribution,$\varepsilon^{1/2}e^{- \varepsilon / k_{\!B}\!T}$, which has already been confirmed by a great deal of MD simulations\cite{21,25,26}. Therefore, within a time unit, the total time for an atom to obtain a kinetic energy larger than the barrier $E_0$($\varepsilon>E_{0}$) is\cite{21}
\begin{equation}
t=\frac{1}{Z}\int_{E_0}^{\infty}\!\varepsilon^{1/2}e^{- \varepsilon / k_{\!B}\!T}d \varepsilon,
\end{equation}
where $Z = \int_{0}^{\infty}\!\varepsilon^{1/2}e^{- \varepsilon / k_{\!B}\!T}d \varepsilon = \sqrt{\pi}(k_{\!B}\!T)^{3/2} /2$
is the partition function.
Considering an atom located at the bottom of a potential well \emph{V}(x)with kinetic energy $\varepsilon$($\varepsilon>E_0$), the time taken by this atom to escape from the valley is $\delta t= \sqrt{m}\int_{0}^{a}dx / \sqrt{2(\varepsilon - V(x))}$, where $a$ is the half width of the well and the average time at a certain temperature $T$ can be obtained by
\begin{equation}
\overline{\delta t}=\frac{\int_{E_0}^{\infty}\!(\delta t)\varepsilon^{1/2}e^{- \varepsilon / k_{\!B}\!T}d \varepsilon}
{\int_{E_0}^{\infty}\!\varepsilon^{1/2}e^{- \varepsilon / k_{\!B}\!T}d \varepsilon}.
\end{equation}
So the frequency (or rate) of the hopping event is
\begin{equation}\label{eqs}
\Gamma = \frac{t}{\overline{\delta t}} = \frac{1}{Z} \cdot \frac{(\int_{E_0}^{\infty}\!\varepsilon^{1/2}e^{- \varepsilon / k_{\!B}\!T}d \varepsilon)^{2}}
{\int_{E_0}^{\infty}\!(\delta t)\varepsilon^{1/2}e^{- \varepsilon / k_{\!B}\!T}d \varepsilon}.
\end{equation}
To apply this model for predicting the stability of adhered MCCs, we should find the key atomic process (with the lowest energy barrier) to break this structure, and calculate the relevant potential energy curve (PEC) along the minimum energy path (MEP). Then we can use Eq.\ref{eqs} to calculate the lifetime ($1/\Gamma$) of this structure at any temperature.

\begin{figure}
\includegraphics[width=8.5cm]{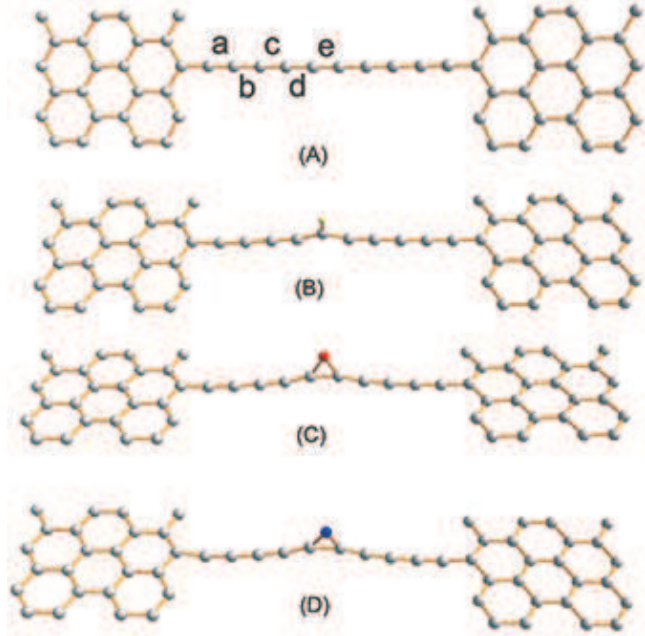}
\caption{\label{structure}(color online). Structures of 10-atom length (A)pristine MCC and MCC with the adhesion of (B)H atom, (C)O atom, and (D)N atom.}
\end{figure}

To get the accurate MEP, first principles calculations were performed using density functional theory (DFT)\cite{28} at the level of Perdew-Burke-Ernzrehof parameterized generalized-gradient approximation\cite{29}. Projector augmented-wave pseudopotentials were used\cite{30} with a cutoff energy of 400 eV. The calculation model is shown in Fig.\ref{structure}, with $3\times3$ cells for each electrode and an MCC between the graphene zigzag edges. K-points sampling of $1\times1\times1$ Monkhorst-Pack grid was employed. The geometry of the system was optimized until the Hellmann-Feynman forces were less than 0.01 eV/{\AA}. To analyze the thermal properties, the potential barrier and the minimum energy path of bond breaking, phonon spectrum and the potential energy surface near the optimized geometry was calculated. The adatom was artificially adhered in all available spots, and then all the structures were optimized. Thus the optimized structures include all the metastable structures in local positions. In the stability calculation process, the precise breakage barrier was obtained at first. And then the life span of the structures can be obtained by our statistical model.

\subsection{Calculations of transport properties}

The models in Fig.\ref{structure} were used to study the properties of MCCs with bilateral semi-infinite graphene electrodes, which are the observed structures in experiments. For quantum transport, calculations were performed using non-equilibrium Green¡¯s function\cite{31,32,33,34}, and the corresponding DFT calculations were performed at the level of Perdew-Burke-Ernzrehof of parameterized generalized-gradient approximation\cite{29} by localized double-zeta polarized basis set and Troullier-Martins pseudopotential\cite{35} with the mesh cutoff energy being 150 Rydberg. The electrode calculations were performed under periodical boundary conditions with k-points sampling of $1\times21\times25$ Monkhorst-Pack grid. For a bias voltage $V_{b}$ applied to the ends of the MCC, the current through the system is given by Landauer-B\"{u}ttiker formula\cite{36},
\begin{equation}
I = \frac{2e}{h}\int T(E,V_{b})[f_{L}(E-E_{F}-\frac{eV_{b}}{2})-f_{R}(E-E_{F}+\frac{eV_{b}}{2})]dE
\end{equation}
where $T(E,V_{b})$ is the scattering coefficient of the band state at energy E, $E_{F}$ is the Fermi energy of the electrodes, $f_{L}$ and $f_{R}$ are Fermi-Dirac distribution functions of both electrodes, respectively.

\section{Results and Discussions}

\subsection{Stability of H, O, and N adhered MCCs}

According to our calculations, the structures of MCC with adhesion of H$_{2}$ and N$_{2}$ molecule own higher energy than the non-adhered structure, so there is no stable adhered structure at room temperature. For O$_{2}$ molecule, when the two oxygen atoms are adhered, the O-O bond will instantly break and form two C-O bond with the neighboring two chain atoms, then making the C-C bond in the middle of the chain break. For these reasons we adhered individual atom (i.e. H, N, and O atom) instead of gas molecules onto MCCs. It is quite obvious that all three different kinds of carbon chains have varying degrees of bending distortion [Fig.\ref{structure} (B,C,D)]. This indicates that new chemical bonds are formed between the adhered atom and chain atoms, as adhered H atom which stays at the top of the carbon atom forms one C-H bond with the neighboring chain atom, while O and N atom alike stay at the bridge site, forming two $\sigma$-bonds with two different chain atoms. The different bonding types make the structural distortion different and H atom makes the largest distortion, as it causes a bending C-C bond angle of nearly $120^{\circ}$, which is sharper than that of O or N atom of about $160^{\circ}$.

The stability of structures after adhesion was mainly determined by the desorption of exotic atom, because it has a lower energy barrier than the breaking of the chain. By MEP, we give the desorption barrier for H, N, and O atom as $2.20, 5.09$ and $4.26$eV, respectively. According to the statistical model [Eq.\ref{eqs}], the lifetime of MCCs consisting of 10 carbon atoms with adhesion of one nitrogen, hydrogen and oxygen atom at room temperature were $10^{5}, 10^{10}$, and $10^{7}$ years, respectively. Therefore in working environments like room temperature, MCCs with atom adhesion are quite stable.

\subsection{Transport Properties of Pristine and H, O and N adhered MCCs}

\begin{figure}
\includegraphics[width=8.5cm]{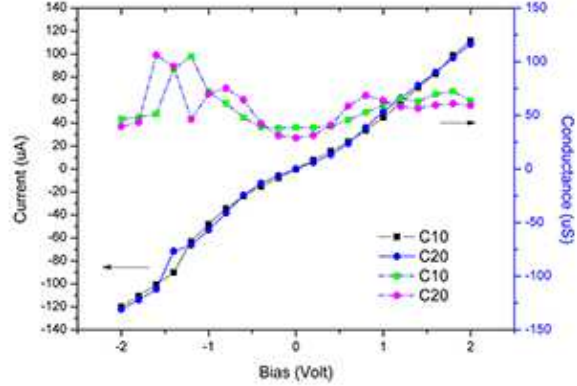}
\caption{\label{CIV}(color online). I-V curves of 10-atom and 20-atom length MCCs. Solid lines correspond to current in units of $\mu$A (left axis), while dash dot lines correspond to conductance in units of $\mu$S (right axis). Square symbol represent C10 MCC and circle symbol represent C20 MCC.}
\end{figure}

The transport properties of pristine MCCs with 10 and 20 chain atoms and I-V curves were shown in Fig.\ref{CIV}. It can be seen that the I-V curves are similar and the conductance is almost independent of the length of the molecular chain. However the differential conductance at zero-bias voltage is $0.50G_{0} (G_{0}= 2e^{2}/h\approx 77.48\mu S)$ for C10 MCC and 0.37$G_{0}$ for C20 MCC, which differs. This result shows the same tendency with the results calculated by Crljen $(1.65G_{0}, 1.56G_{0}, 1.49G_{0}$, and 1.44$G_{0}$ for C2, C4, C6, and C8, respectively\cite{15}) that the differential conductance of even numbered MCCs decreases as the chain length increasing. The smaller conductance in the present work may be resulted from the use of graphene electrodes compared with gold electrodes used in Ref.5.

\begin{figure}
\includegraphics[width=8.5cm]{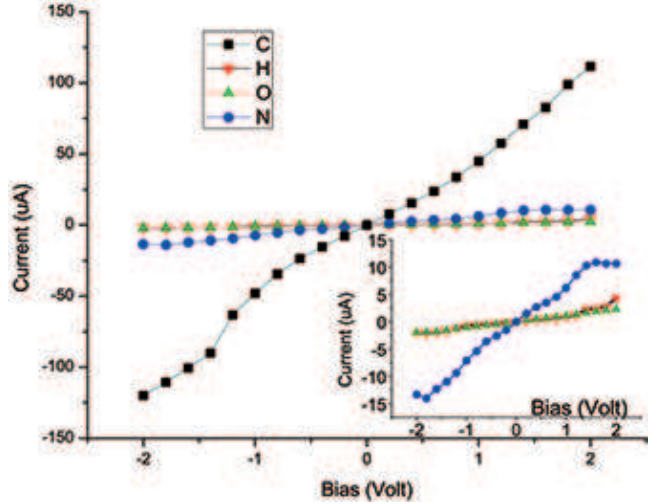}
\caption{\label{IV}(color online). I-V curves of pristine MCC (square) and MCCs with the adhesion of H atom(inverted triangle), O atom(triangle) and N atom(circle). The inset shows the enlarged I-V curves of MCCs with adhesion. Square represent for the adhesion of N atom, circle for H atom and triangle for O atom.}
\end{figure}

Fig.\ref{IV} shows the I-V curves of a 10-atom length pristine MCC and MCC with adhesion of H, O, and N atom. From the I-V curves, we see that the conductance of MCC is reduced by one to two orders of magnitude after adhesion, and the inset shows that the MCC with the adhesion of N atom own larger conductance than that of H or O atom and that different atoms have a different impact on conductance. It suggests that the adhesion of exotic atom indeed can change the conductance of MCC largely.

\begin{figure}
\includegraphics[width=8.5cm]{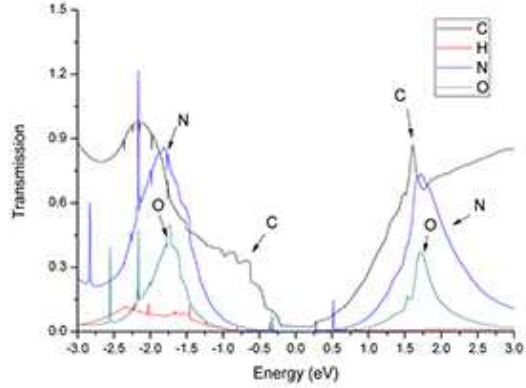}
\caption{\label{Trans}(color online). Transmission spectra of pristine MCC, and MCCs with adhesion of H atom, N atom and O atom under zero-bias. The x-axis shows the energy measured with respect to the Fermi level. The transmission spectrum for pristine MCC is in black, MCC with adhesion of H atom in red, N atom in blue and O atom in green.}
\end{figure}

To understand the large conductance decrease of MCCs by the adhered atoms, the transmission spectra for the pristine and H, N and O adhered MCCs were calculated under zero bias and were shown in Fig.\ref{Trans}. As the electron transport in this system is mainly in a tunneling picture, i.e. as molecular energy level of MCCs resonates with the incident energy of the electron from the electrode, one transport channel opens, which corresponds to a peak in the transmission spectra, so the transmission peak in Fig.\ref{Trans} should correspond to the molecular energy level of MCCs. Compared with pristine MCCs, the energy level of adhered MCCs barely changes, and the main difference is that the amplitude of transmission peaks decreases much [Fig.\ref{Trans}], which leads to the decrease of conductance. This change is especially obvious for H atom, where the transmission peaks are nearly suppressed near the Fermi level, so its conductance is the smallest compared with N and O atom. To explain the transmission decrease, we resort to the distorted structures of MCCs after adhesion. As the electron transport is along the chain axis, the structural distortion at the adhesion site is certain to scatter the electron, and it can be expected that the more distorted structure will cause stronger scattering thus decrease the transmission. This analysis is confirmed by comparing the distortions between H, N and O, that the C-C bond angle at the adhesion site is sharpest for adhered H atom [Fig.\ref{structure}], and the transmission amplitude of MCC with the adhesion of H atom is indeed the smallest [Fig.\ref{Trans}].

\begin{figure}
\includegraphics[width=8.5cm]{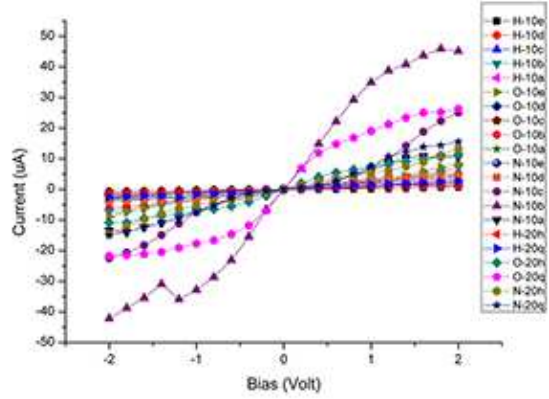}
\caption{\label{AllIV}(color online). I-V curves of 10-atom and 20-atom length pristine MCC and MCCs with adhesion of H, O, and N atom on all available sites.}
\end{figure}

Then we focused on the influence of adhesion site on the transport properties of MCCs. All the five possible adhesion sites denoted by $a, b, c, d$ and $e$ in the chains [Fig.\ref{structure}A] were tested by adhering one H, N or O atom, and the calculated I-V curves are shown in Fig.\ref{AllIV}. It can be seen that the conductance is not only sensitive to the adhesion atom species but also to the adhesion site. For example, the conductance of N adhered chain on $b$ site is nearly 50 times of that on $e$ site. No clear tendency was found when considering how the adhesion site influence the conductance, as the largest conductance of N adhered chain comes from the adhesion on b site, but that comes from e site for O adhered chain. On average, the MCCs with the adhesion of N atom own larger conductance than that of H and O atoms, which are comparable with many adhesion sites.

\begin{figure}
\includegraphics[width=8.5cm]{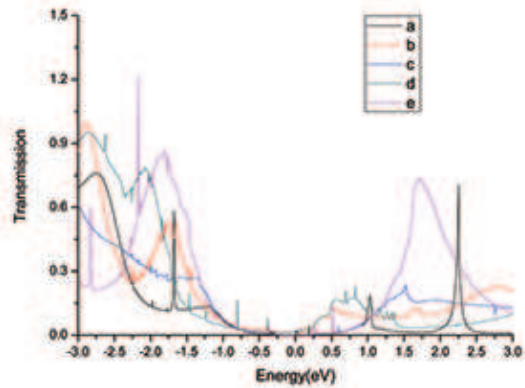}
\caption{\label{Ntrans}(color online). Transmission spectra of MCCs with adhesion of N atoms under zero-bias on five adhesion sites. The x-axis shows the energy measured with respect to the Fermi level.}
\end{figure}

It is notable that the structural distortion of the chain barely changes as one atom is adhered on different sites, thus the various conductance changes cannot be explained simply by the structural distortions. To further investigate the conductance change, we take the adhesion of N atom as an example to calculate the transmission spectra on different adhesion sites [Fig.\ref{Ntrans}]. The transmission spectra seem quite different for different adhesion sites, as both the number of peaks and their energy change, which indicates the Fermi energy and the energy levels of MCCs change with the different adhesion sites. This fact can be conventionally accepted because the MCC in this calculation consists only 10 atoms, and chemically bonding one atom to this system may have a great influence on the chain's electronic structure.

\section{Conclusion}

In summary, ways to tune the conductance of MCCs by adhering H, N and O atoms to the chains were investigated. By a statistical model, we first showed that the structures after adhesion are quite stable at room temperature. Then by transport calculation, we found that the conductance of MCCs decreases by one to two orders of magnitude after adhesion. The conductance of the adhered MCCs changes largely with different adhesion atom species and adhesion sites, and no simple relationship between the conductance and adhesion atom species or sites was found. By analyzing the transmission spectra, we attributed the conductance change to the chemical bonding between the adhesion atom and chain atoms, which causes distortion of the chain structure and forms new molecular energy levels.

\begin{acknowledgments}
The authors are very grateful to acknowledge Professor Qike Zheng for helpful discussions. This work was supported by the National Natural Science Foundation of China under Grant No. 11274073, 11074042 and 51071048, Shanghai Leading Academic Discipline Project (Project No. B107) and Key Discipline Innovative Training Program of Fudan University.
\end{acknowledgments}



%

\end{document}